\documentclass[prl,tightenlines, twocolumn,letterpaper,nofootinbib,showpacs]{revtex4}
\usepackage{epsfig,amsmath,amssymb,bm,epsf,graphicx,latexsym,amsfonts,hyperref}

\def\ket#1{ | #1 \rangle}
\def\bra#1{{\langle #1 | }}
\def\tr{{\rm{Tr}}}
\def\vec#1{\boldsymbol{#1}}

\begin{document}

\title{Doubly constrained bounds on the entanglement of formation}

\author{Animesh Datta}
\email{animesh@unm.edu}
\author{Steven T. Flammia}
\author{Anil Shaji}
\author{Carlton M. Caves}
\affiliation{Department of Physics and Astronomy, University of New
Mexico, Albuquerque, New Mexico 87131-1156}


\begin{abstract}
We derive bounds on the entanglement of formation of states of a $4\times
N$ bipartite system using two entanglement monotones constructed from
operational separability criteria.  The bounds are used simultaneously as
constraints on the entanglement of formation.  One monotone is the
negativity, which is based on the Peres positive-partial-transpose
criterion.  For the other, we formulate a monotone based on a separability
criterion introduced by Breuer (H.-P. Breuer, e-print quant-ph/0605036).
\end{abstract}



\maketitle

The nonclassical correlations of entangled quantum states~\cite{brub02a}
have been of interest since the very inception of quantum
mechanics~\cite{einstein35a,schrodinger35c}.  Quantum information science
has led to the idea that entanglement is a resource for information
processing and other tasks.  The ability of quantum computers to solve
classically hard problems efficiently, the increased security of quantum
cryptographic protocols, and the enhanced capacity of quantum
channels---all these are attributed to entanglement. Investigating
entanglement has led to new understanding of techniques such as the
density-matrix renormalization group~\cite{vidal03a} and of quantum phase
transitions~\cite{on02,osterloh02a} and properties of condensed
systems~\cite{grac03}.  Despite the importance of entanglement, however,
characterizing and quantifying it in most physical systems that are of
interest from an experimental standpoint remains a challenge.

An important measure of entanglement for a pure state $\ket{\Psi}$ of two
systems, $A$ and $B$, is the entropy, $-\tr(\rho_{A}\log\rho_{A})$, of the
marginal density operator $\rho_A$.  We write this entropy sometimes as a
function $h(\Psi)$ and sometimes as the Shannon entropy $H(\vec\mu)$ of
the vector $\vec\mu$ of Schmidt coefficients of $\ket{\Psi}$.  This
measure can be applied to bipartite mixed states by the convex-roof
extension of $h(\Psi)$. The extended quantity, called the {\em
entanglement of formation\/} (EOF), is defined~as
\begin{equation}
\label{eof} h(\rho)\equiv\!
\min_{\{p_j,\ket{\Psi^j}\}}\!
\bigg\{\!\sum_j p_j h(\Psi^j)\,\bigg|\,\rho=\sum_jp_j\ket{\Psi^j}\bra{\Psi^j}\bigg\}\;.
 \end{equation}
The EOF is a {\em nonoperational\/} measure of entanglement because the
minimization over all pure-state decompositions of $\rho$ generally means
there is no efficient procedure for calculating it.  This minimization is
the bottleneck in evaluating most nonoperational entanglement measures for
mixed states.  Consequently, bounding the EOF, instead of computing its
value, becomes important.

An alternate approach to quantifying entanglement is based on the use of
positive (but not completely positive) maps on density
operators~\cite{choi72}.  A quantum state is {\em separable\/} if and only
if it remains positive semidefinite under the action of {\em any\/}
positive map.  Given a positive map, we can construct a related
entanglement monotone by considering the spectrum of density operators
under the action of the map~\cite{plenio05a,vidal02a}.  Such monotones are
typically much easier to calculate for general quantum states, because
they do not involve the convex-roof construction, and thus are said to be
{\em operational\/}~\cite{brub02a}.

We can use the  monotones constructed from positive maps and from other
operational entanglement criteria as constraints to obtain bounds on
nonoperational, convex-roof extended measures of entanglement.  The
complexity of the minimization in Eq.~(\ref{eof}) is reduced by solving it
over a constrained set, instead of over all pure-state decompositions.
This was done in~\cite{terhal00a,chen05a} for the EOF, using a single
operational constraint.  Our endeavor in this Letter is to carry this
program forward. We first sketch a general scheme for many constraints,
which we discuss further in~\cite{datta06a}, and then illustrate the
general scheme for a particular case of two operational constraints.

Let us say that $f_1,\ldots,f_K$ are operational entanglement monotones
for a bipartite system.  We gather their values for an arbitrary state
$\rho $ into a vector $\mathbf{n} = (n_1,\ldots,n_K)$.  Their actions on
pure states are functions of the Schmidt coefficients, i.e., $f_k(\Psi)
=F_k(\vec{\mu})$ for $k=1,\ldots,K$.

We are interested in a lower bound on the value of the {EOF}. Let us
assume that for the state $\rho$, the optimal pure-state decomposition is
$\rho = \sum_j p_j |\Psi^j \rangle \langle \Psi^j|$, giving $h(\rho) =
\sum_j p_j H\bigl(\vec{\mu}^j\bigr)$. Now define the function
\begin{equation}
\label{eq:doublyA3}
 \widetilde{H}({\bf m}) = \min_{\vec{\mu}}
 \Bigl\{H(\vec{\mu})\!\Bigm|\!F_k(\vec{\mu})=m_k,\;k=1,\ldots,K\Bigr\}\;.
\end{equation}
Notice that $\widetilde{H}({\bf m})$ is defined only on the region of
possible values of ${\bf m}$ corresponding to pure states, a region we
call the {\em pure-state region}.  If ${\widetilde H}$ is not a
monotonically nondecreasing function of ${\bf m}$, which we will call a
monotonic function for brevity, we replace it with such a monotonic
function $\widetilde{H}_{\uparrow}({\bf m})$, constructed by dividing the
pure-state region into subregions on which subsets of the constraints are
applied.  We describe the procedure for constructing
$\widetilde{H}_{\uparrow}({\bf m})$ in detail in~\cite{datta06a}.

Let ${\mathcal H}({\bf m}) = {\mbox{co}} [ \widetilde{H}_\uparrow ({\bf
m})]$ be the convex hull of $\widetilde{H}_\uparrow({\bf m})$, i.e., the
largest convex function of $K$ variables $(m_1,\ldots,m_K)$ bounded from
above by $\widetilde{H}_\uparrow({\bf m})$.  We can show that $\mathcal
H({\bf m})$ is also a montonic function~\cite{datta06a}, which can be
extended naturally to a monotonic function on the entire space of values
of ${\bf m}$.  Using Eq.~(\ref{eq:doublyA3}) and the convexity and
monotonicity of ${\mathcal H}$, we can write
\begin{equation}
\label{eq:doublyA6}
h(\rho) \geq
\sum_j p_j{\mathcal H}({\bf n}^j)
\geq {\mathcal H} \bigg( \sum_j p_j {\bf n}^j \bigg)
\geq {\mathcal H} ({\bf n})\;,
\end{equation}
where we have used the convexity of the monotones $f_k$ to obtain $\sum_j
p_j n_k^j \geq n_k$.  Knowing the easily calculated $\mathbf{n}$ for
$\rho$ thus leads to a bound on $h(\rho)$.

We now carry through the general program for $4\times N$ states using two
operational entanglement monotones as constraints.  Ours is the first
instance of a doubly constrained bound on an entanglement measure for a
family of states.  It gives tighter bounds than those obtained
previously~\cite{chen05a}.

The first monotone is the {\em negativity\/}~\cite{vidal02a}, which is
based on the Peres criterion~\cite{peres96a}. The negativity of a
bipartite state $\rho$ is defined as $ n_T(\rho) = (||\rho^{T_A}||-1)/2$
where $T_A$ is the partial transposition with respect to system $A$ and
the trace norm is defined as $||O||=\tr(\sqrt{OO^{\dagger}})$. For pure
states, the negativity, in terms of the Schmidt coefficients, is given by
$n_T = [( \sum_j \sqrt{\mu_j} )^2-1]/2$.

We define a second monotone based on the $\Phi$-map introduced by
Breuer~\cite{breuer06a}. The action of the $\Phi$-map on any state
$\sigma$ is given by $\Phi(\sigma) = \tr(\sigma)I -\sigma - V\sigma^T
V^{\dagger}$, where the superscript $T$ stands for transposition and $V$
is a unitary matrix with matrix elements $\bra{j,m}V\ket{j,m'}=
(-1)^{j-m}\delta_{m,-m'}$ in the angular momentum basis $\{ | j,m\rangle
\}$. The map $\Phi$ provides, for any bipartite state $\rho$ having a
subsystem with even dimension greater than 4, a nontrivial condition for
separability as $(I\otimes\Phi)(\rho)~\geq~0$.  The related entanglement
monotone, which we call the {\em $\Phi$-negativity}, is defined for a
general mixed state as
 \begin{equation}
 \label{nphi}
 n_{\Phi}(\rho)=\frac{D(D-1)}{4}\left[\frac{||(I\otimes\Phi)(\rho)||}{D-2}-1\right]\;,
 \end{equation}
where $D$ is the dimension of the smaller of the two systems in the
bipartite state $\rho$. The $\Phi$-negativity is a convex function of
$\rho$. For $4 \times N$ systems $(N \geq 4)$, the $\Phi$-negativity for
pure states, as a function of the four Schmidt coefficients, is
$n_{\Phi}=3\sqrt{(\mu_1+\mu_4)(\mu_2+\mu_3)}$.  The $\Phi$-negativity for
various states is given in~\cite{datta06a}.

We can place bounds on the EOF of $4 \times N$ states by using either
$n_{\Phi}$ or $n_T$ as constraints.  To find the bound with $n_T$ as the
single constraint, which was done in~\cite{chen05a}, one first finds the
singly constrained function $\widetilde{H}(n_T$) of
Eq.~(\ref{eq:doublyA3}).  This function being monotonic, but not convex,
its convex roof gives the bound.  For the $4\times N$ states we consider,
the bound is given by
 \begin{equation}
\label{Tbound}
\begin{array}{l}
 {\mathcal H}\left( n_T \right)=
 \left\{ \begin{array}{ll}
 H_{2}(\gamma)+(1-\gamma)\log _{2}3\;, & n_T \in [0,1],\\[2mm]
 \big(n_T -\frac{3}{2}\big)\log _2 3+ 2\;, & n_T \in [1,\frac{3}{2}],
\end{array} \right.
\end{array}
\end{equation}
where $H_2$ is the binary entropy function and $\gamma=(\sqrt{2n_T+1} +
\sqrt{9-6n_T}\,)^2/16$.  If instead we use $n_{\Phi}$ as the single
constraint, we first find the function $\widetilde{H}(n_\Phi)$, which
being monotonic and convex, gives directly a different bound on the EOF of
$4 \times N$ states~\cite{datta06a},
\begin{equation}
 \widetilde{H}(n_\Phi)={\mathcal H}\left( n_{\Phi} \right)
 = H_2(\alpha)\;, \quad
 \alpha =\frac{1+\sqrt{1-4 n_\Phi^2/9}}{2}\;.
 \end{equation}
We refer to ${\mathcal H}\left( n_{\Phi} \right)$ and ${\mathcal H}\left(
n_T \right)$ as {\em singly constrained}\/ bounds on the {EOF}.  We now
proceed to place a {\em doubly constrained bound}\/ on the EOF of $4 \times
N$ density operators by simultaneously using $n_T$ and $n_{\Phi}$ as
constraints.

Both $n_{\Phi}$ and $n_T$  take on values between $0$ and $3/2$, so all $4
\times N$ states lie in a square of side $3/2$ in the $n_{\Phi}$-$n_T$
plane.  Not all points in the square correspond to pure states.  Solving
simultaneously the normalization constraint $\sum_{j=1}^4 \mu_j=1$ and the
two constraint equations, $\sum_{j=1}^4 \sqrt{\mu_j} = \sqrt{2n_T+1}$ and
$3 \sqrt{(\mu_1 + \mu_4)(\mu_2 + \mu_3)} = n_{\Phi}$, lets us express
$\mu_1$, $\mu_2$, and $\mu_3$ in terms of $n_{\Phi}$, $n_T$, and $\mu_4$.
For some values of $n_{\Phi}$ and $n_T$, there is no value for $\mu_4$ for
which the other three Schmidt coefficients are real numbers in the
interval $[0,1]$.

\begin{figure}
\begin{center}
\resizebox{6 cm}{6 cm}{\includegraphics{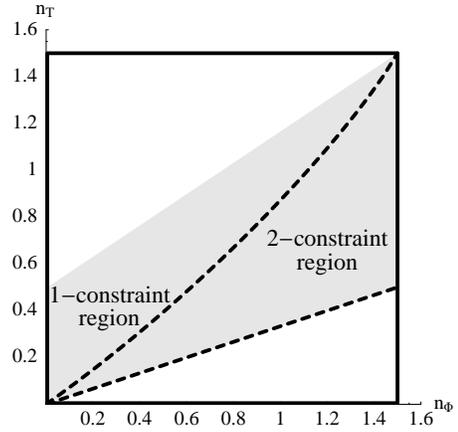}}
 \caption{The shaded region is the pure-state region in the
 $n_{\Phi}$-$n_T$ plane for $4\times N$ pure states.  The dashed lines are
 the monotone boundaries given by Eq.~(\ref{eq:doublyC2}) and by $n_T =
 n_\Phi/3$.  In the $2$-constraint region between the monotone boundaries,
 we set $\widetilde{H}_\uparrow({\bf n})= \widetilde{H}({\bf n})$, and in
 the 1-constraint region above the upper monotone boundary, we set
 $\widetilde{H}_\uparrow({\bf n})= \widetilde{H}(n_T)$.}
 \label{fig:pure_regions}
\end{center} \vspace{-9 mm}
\end{figure}

To find the pure-state region, we look for the maximum and minimum allowed
values of $n_T$ for a fixed $n_\Phi$, assuming a pure state.  To find the
maximum, we apply the technique of Lagrange multipliers and obtain $n_T =
2 n_{\Phi}/3 + 1/2$.  The minimum lies on the boundary of allowed values
of $\vec{\mu}$, with $\mu_3=\mu_4=0$, and is given by $n_T= n_\Phi/3$.
The resulting pure-state region, shown in Fig.~\ref{fig:pure_regions}, is
convex.  The pure-state region is not convex in general, however; the
subtleties this introduces into our program are addressed
in~\cite{datta06a}.

To find the doubly constrained bound on the EOF, we start with the
function~(\ref{eq:doublyA3}), specialized to our two constraints,
\begin{eqnarray}
\label{eq:doublyC1}
\widetilde{H}(n_{\Phi},n_T )\!\! &\equiv&
\!\!\min_{\vec{\mu}} \bigg\{ H(\vec{\mu}) \bigg|  \sum_j \sqrt{\mu_j}
=\sqrt{2 n_T+1},  \nonumber \\
&& \hspace{.7 cm} 3 \sqrt{(\mu_1+ \mu_4)(\mu_2+\mu_3)} = n_{\Phi} \bigg\}\;.
\quad
\end{eqnarray}
It turns out that $\widetilde{H}(n_{\Phi},n_T )$ is not monotonic, so we
must replace it with the monotonic function
$\widetilde{H}_\uparrow(n_{\Phi},n_T )$ discussed above.  The procedure
for constructing $\widetilde{H}_\uparrow(n_{\Phi},n_T )$, depicted in
Fig.~\ref{fig:pure_regions}, makes a connection to the singly constrained
bounds.  This connection is based on the fact that the minimum of {\em
any}\/ function subject to two constraints is greater than or equal to the
minimum of the same function subject to only one of the two constraints.
Thus we can say that $\widetilde{H}(n_\Phi,n_T)\ge\widetilde{H}(n_T)$ for
all $n_\Phi$ and $\widetilde{H}(n_\Phi,n_T)\ge\widetilde{H}(n_\Phi)$ for
all $n_T$.

The minimum of $H(\vec{\mu})$ subject only to the $n_T$ constraint, i.e.,
$\widetilde{H}(n_T)$, occurs when the Schmidt coefficients are given by
$\vec{\mu}=(\gamma, \gamma', \gamma', \gamma')$~\cite{chen05a} with
$\gamma'=(1-\gamma)/3$.  This corresponds to $n_{\Phi}=
\sqrt{2(2\gamma+1)(1-\gamma)}$, thus defining a curve in the
$n_\Phi$-$n_T$ plane.  Writing $\gamma$ in terms of $n_T$ puts this curve
in the form
\begin{equation}
\label{eq:doublyC2} n_T \! =\!
\frac{3}{4} \!\! \left( \! 1 \!- \! \sqrt{1
- \frac{4}{9} n_{\Phi}^2}\! + \!\sqrt{\frac{4}{3} n_{\Phi}^2 \! + \!2
\sqrt{ \! 1 \!- \! \frac{4}{9} n_{\Phi}^2} \!  - \! 2}\,\right)\;.
\end{equation}
Along this curve, which we call a {\em monotone boundary}, the $n_\Phi$
constraint is automatically satisfied when $H(\vec\mu)$ is minimized with
respect just to the $n_T$ constraint, which means that
$\widetilde{H}(n_\Phi,n_T)=\widetilde H(n_T)$ on this monotone boundary.
To construct the required monotonic function, we set
$\widetilde{H}_\uparrow(n_\Phi,n_T)= \widetilde{H}(n_T)$ when
$n_\Phi\le\sqrt{2(2\gamma+1)(1-\gamma)}$, i.e., above this monotone
boundary.

Similarly, the minimum of $H(\vec\mu)$ subject just to the $n_{\Phi}$
constraint, i.e., $\widetilde{H}(n_\Phi)$, occurs when
$\vec{\mu}_{\Phi}=(\alpha,1-\alpha,0,0)$, which gives a lower monotone
boundary $n_T= \sqrt{\alpha (1-\alpha)}=n_{\Phi}/3$.  Along this line, the
$n_T$ constraint is automatically satisfied when $H(\vec\mu)$ is minimized
with respect just to the $n_\Phi$ constraint, which gives
$\widetilde{H}(n_\Phi,n_T)=\widetilde H(n_\Phi)$ on this boundary.  Since
this lower monotone boundary coincides with the lower boundary of the
pure-state region, it has no impact on defining
$\widetilde{H}_\uparrow(n_\Phi,n_T)$.

The definition of $\widetilde{H}_\uparrow({\bf n})$ is depicted in
Fig.~\ref{fig:pure_regions}.  Between the monotone boundaries, a region we
call the {\em $2$-constraint region}, we set $\widetilde{H}_\uparrow({\bf
n})= \widetilde{H}({\bf n})$, and in the pure-state region above the upper
monotone boundary, which we call the {\em $1$-constraint region}, we set
$\widetilde{H}_\uparrow({\bf n})= \widetilde{H}(n_T)$.  The resulting
function $\widetilde{H}_\uparrow({\bf n})$ is monotonic throughout the
pure-state region.

We now focus on finding $\widetilde{H}({\bf n})$ in the $2$-constraint
region.  The method of Lagrange multipliers is not suitable for finding
the minimum~(\ref{eq:doublyC1}) because the problem is overconstrained.
The equations obtained using Lagrange multipliers have a consistent
solution only if $n_{\Phi}$ and $n_T$ are related as in
Eq.~(\ref{eq:doublyC2}), in which case $\widetilde{H}({\bf
n})=\widetilde{H}(n_T)$.  This does not mean that there is no minimum for
$H(\vec{\mu})$ for other values of $n_\Phi$ and $n_T$, just that the
minimum lies on a boundary of allowed values of $\vec{\mu}$.  The boundary
with three of the Schmidt coefficients being zero is the origin in the
$n_{\Phi}$-$n_T$ plane, where $H(\vec{\mu})=0$.  The boundary with two
zero Schmidt coefficients is the line $n_T=n_{\Phi}/3$, and along this
line $\widetilde{H}({\bf n})=\widetilde{H}( n_{\Phi})$.

The minimum of $H(\vec{\mu})$ in the remaining part of the $2$-constraint
region can be found using a straightforward numerical procedure.  As
discussed above, the constraint equations can be solved to express
$\mu_1$, $\mu_2$, and $\mu_3$ in terms of $n_{\Phi}$, $n_T$, and $\mu_4$.
There are two distinct solutions, $\vec{\mu}^{(1)}$ and $\vec{\mu}^{(2)}$.
For a particular value of $\mu_4$, one or both of these solutions can be
invalid in parts of the pure-state region because one or more of the three
Schmidt coefficients lies outside the interval $[0,1]$.  For valid
solutions we compute the entropy $H(\vec\mu)$.

\begin{figure}[!ht]
\begin{center}
\resizebox{6 cm}{6 cm}{\includegraphics{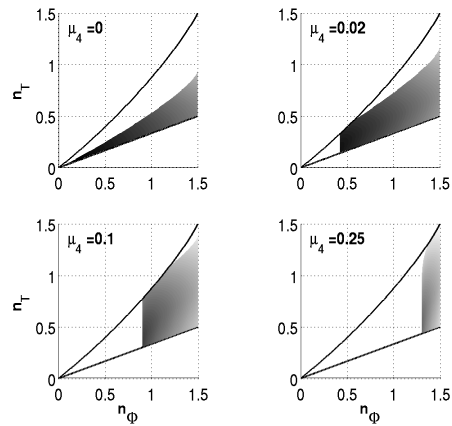}}
 \caption{The part of the $2$-constraint region covered by four values of $\mu_4$.
 The two lines are the monotone boundaries.}
 \label{fig:m4regions}
\end{center} \vspace{-7 mm}
\end{figure}

We first consider the boundary where one Schmidt coefficient is zero by
setting $\mu_4=0$ in the solutions $\vec{\mu}^{(1)}$ and
$\vec{\mu}^{(2)}$.  Not all points in the $2$-constraint region can be
reached if we set $\mu_4=0$.  This is easily seen by noticing that the
point $n_{\Phi}=n_T=3/2$ corresponds uniquely to a maximally entangled $4
\times N$ state, and for this state all four Schmidt coefficients have the
value $1/4$.  Indeed, a continuum of points cannot be reached if we stay
on the boundary defined by $\mu_4=0$, so we increase the value of $\mu_4$
in small steps.  The parts of the $2$-constraint region that are covered
by four values of $\mu_4$ are shown in Fig.~\ref{fig:m4regions}.

This numerical procedure gives us, for each point ${\bf n}=(n_\Phi,n_T)$
in the pure-state region, the range of values of $\mu_4$ for which
$H\big(\vec \mu^{(1)}\big)$ and/or $H\big(\vec \mu^{(2)}\big)$ can be
calculated at that point.   The minimum of these entropies over the
allowed range of values for $\mu_4$ is the value of $\widetilde{H}({\bf
n})$.

The function $\widetilde{H}({\bf n})$ in the $2$-constraint region is, as
required, a monotonic function of both $n_{\Phi}$ and $n_{T}$.  It is
extended to the monotonic function $\widetilde{H}_\uparrow({\bf n})$ on
the the entire pure-state region using the procedure outlined above.  The
monotonic function $\widetilde{H}_{\uparrow}({\bf n})$ is not convex,
however, so we must compute its convex hull ${\mathcal H}({\bf n})$.  This
can be done numerically, and it turns out that the difference between
${\mathcal H}({\bf n})$ and $\widetilde{H}_{\uparrow}({\bf n})$ is quite
small ($\sim 10^{-3}$), the two functions differing only in a small area
near the maximally entangled state.  Had the pure-state region, on which
$\widetilde{H}_{\uparrow}({\bf n})$ is defined, not been convex,
${\mathcal H}({\bf n})$ would be defined on an extended convex
domain~\cite{datta06a}.

\begin{figure}
\begin{center}
\resizebox{6 cm}{10 cm}{\includegraphics{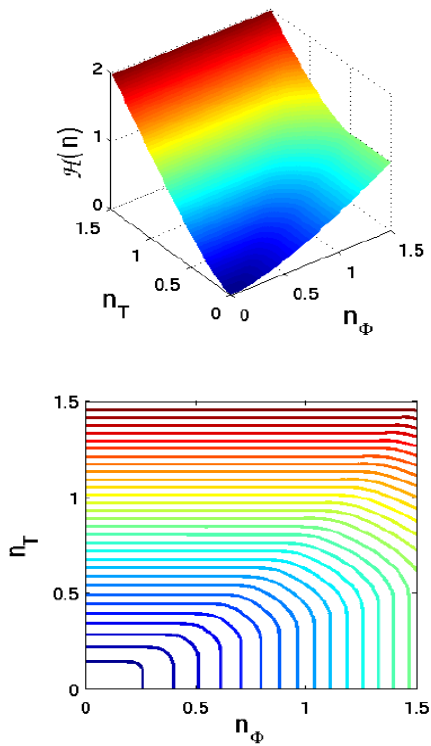}}
 \caption{(Color online) The doubly constrained bound ${\mathcal H}({\bf
 n})$ on the EOF of $4 \times N$ states.  Also shown is a contour plot of
 the same function.} \label{fig:eofbound}
\end{center} \vspace{-10 mm}
\end{figure}

To obtain a bound on the EOF of {\em all\/} $4 \times N$ states, we have
to extend ${\mathcal H}({\bf n})$ outside the pure-state region to the
rest of the $n_{\Phi}$-$n_T$ plane.  The extension has to preserve the
monotonicity of ${\mathcal H}({\bf n})$ so that the string of inequalities
in Eq.~(\ref{eq:doublyA6}) holds.  This is achieved by extending
${\mathcal H}({\bf n})$ using surfaces that match the function at the
lower and upper boundaries of the pure-state region. To preserve
monotonicity, the surface added in the region below the lower boundary has
zero slope along the $n_T$ direction, and the surface added in the region
above the upper boundary has zero slope along the $n_{\Phi}$ direction.
The resulting doubly constrained bound ${\mathcal H}({\bf n})$ on the EOF
is shown in Fig. \ref{fig:eofbound}.  The figure indicates that the
extension to the whole $n_{\Phi}$-$n_T$ plane produces a smooth and
seamless surface.

A third constraint based on the realignment
criterion~\cite{rudolph02a,chen03a} can be used to improve our bound on
the EOF for certain classes of states.  We can define the realignment
negativity for a bipartite density operator $\rho$ as $n_R = (||{\mathcal
R}(\rho)|| -1)/2$, where $\left[{\mathcal R}(\rho)\right]_{ij,kl} =
\rho_{ik,jl}$.  For pure states, $n_R=n_T$. This means that in deriving
the bounds, we could have redefined $n_T$ as $\max(n_T, n_R)$.

In this Letter we focused on the derivation of a particular doubly
constrained bound on the EOF of $4\times N$ systems.  Starting from the
$\Phi$-map introduced by Breuer~\cite{breuer06a,breuer06b}, we defined an
entanglement monotone, the $\Phi$-negativity, and combined it with the
usual negativity to formulate a doubly constrained bound.  We found that
the pure-state region in the $n_\Phi$-$n_T$ plane is divided into sectors
by monotone boundaries.  The doubly constrained pure-state marginal
entropy is applicable only in the region between the monotone boundaries.
In the remaining portions of pure-state region, singly constrained
entropies are applicable.  Monotonicity and convexity dictate how to
extend the bound to all states.  We expect these features to persist for
systems that are not $4\times N$ and for more than two constraints, in
which case the monotone boundaries will generally be hypersurfaces.  A
sector in which an $m$-constrained marginal entropy holds will be bounded
by sectors in which $(m-1)$-constrained marginal entropies hold.  These
methods might provide a useful procedure for bounding the EOF and other
convex-roof entanglement monotones.

This work was supported in part by Office of Naval Research grant No.~N00014-03-1-0426.

\vspace{-5 mm}

\bibliography{eof}

\end{document}